\documentclass[12pt,thmsa]{article}
\usepackage{amssymb}
\usepackage{graphicx}

\makeatletter
\newcommand{\row}[1]{\mathord{\buildrel{\lower3pt\hbox{$\scriptscriptstyle\rightarrow$}}\over #1}}

\newcommand{\dyadic}[1]{\mathord{\dyadic@rrow{#1}}}
\newcommand{\dyadic@rrow}[1]{
\begin{picture}(12,12)(-1,0)
\put(-2,10){\makebox(0,0)[t]{$\scriptscriptstyle\downarrow$}}
\put(-2,11){\makebox(0,0)[l]{$\scriptscriptstyle\longrightarrow$}}
\put(5,0){\makebox(0,0)[b]{$#1$}}
\end{picture}
}
\newcommand{\bra}[1]{\bigl\langle #1 \bigr|}
\newcommand{\ket}[1]{\bigl| #1 \bigr\rangle}
\newcommand{\expect}[1]{\left\langle #1 \right\rangle}

\input{tcilatex}
\begin{document}

\begin{center}
\textbf{Entanglement and teleportation via chaotic system }

N. Metwally$^{1}$, L. Chotorlishvili$^{2,3}$, V. Skrinnikov$^{3}$ \\[0pt]

$^{1}$ Math. Dept., College of Science, Bahrain University, 32038 Bahrain.
\\[0pt]

$^{2}$Institute for Physik, Universit$\ddot{a}$t Augsburg, 86135  Augsburg,
Germany

$^{3}$Physics Department of the Tbilisi State University, Chavchavadze av.3,
0128, Tbilisi
\end{center}

\textbf{{Abstract}: } The dynamics of entangled state interacting
with a single cavity mode is investigated in the presence of a
random parameter. We have shown that degree of entanglement decays
with time and rate of decay is defined by features of random
parameter. Quantum teleportation  through dissipative channal and
teleportation fidelity as a function of damping rates has been
studied. The sensitivity of the fidelity with respect to random
parameter is discussed. We have evaluated the time interval during
which  one can  perform the quantum teleportation and send the
information with reasonable fidelity, for a given values of
correlation length of random parameter.


\section{Introduction}

Entanglement is a promising resource for the quantum information and
communication fields \cite{Niel}. There are much attentions have
been paid to the manipulation of quantum entanglement\cite{Peres}.
Quantum teleportation \cite{ben} and quantum
cryptography\cite{Ekert,ben2} all rely on having entangled states of
qubits. Therefore generating entangled states and
keeping them survival for along time is an important task \cite%
{khasin}. In the real word, quantum entanglement will be inevitably
affected by the deocherence, which causes deterioration of entanglement \cite%
{Wang}. Thus, there are several studies which concentrate on
investigating the dynamics of quantum entanglement and information
in open systems (see for example \cite{Dubi,Isar}). Also, the
thermal effect on the dynamic of the information has been treated in
many different systems. As an example, the effect of the local
thermal and squeezed reservoirs on the degree of entanglement is
investigated by Ikram et. al \cite{Ikram}. The entanglement decay
and the sudden death in the presences of thermal reservoir is
discussed by Zhou and et. al \cite{Zhou}. The dynamics of quantum
information is investigated by means of the information exchange and
the disturbance in \cite{nas}.

Entanglement  affected by other factors than  thermal reservoirs.
For example, system may contain stochastic parameter averaging with
respect to which leads to the de-coherence. Problems like this are
typical for Cavity Quantum Electrodynamics (CQED). Modern
experiments in CQED have achieved strong atom-field coupling for the
strength of the coupling exceeds decay process. If so, then dynamic
of two level atoms inside a cavity is reduced to the driven
Jaynes-Cummings (JC) Hamiltonian, which models the interaction of a
single mode of an optical cavity having resonant frequency with a
two level atom \cite{Schleich,Bin}. However as was shown in
\cite{Zaslavsky}, in  the most general case dynamics is very
complicated and chaotic. Since it involves not only the internal
atomic transitions and field states, but also center-of-mass motion
of the atom and recoil effect. Motion of two level optical atoms
inside a cavity can be chaotic \cite{Zaslavsky}. So, the  coordinate
of the atom is a random parameter and leads to the sufficient
influence on the quantum information processing.

It is well-known that, in a perfect scheme, the shared entangled
state is a maximally entangled state enabling perfect quantum state
transfer. However interaction with the environment results in a loss
of coherence. This phenomenon is key issue of many studies
\cite{Dubi,Isar}). However, the main idea of the given paper is to
study the entangled states in the presence of inner random parameter
in the system. For this purpose, we shall consider a generalized JC
model introduced in \cite{Levan}. Also, the possibility of
performing quantum teleportation through this chaotic system will be
discussed. Study of the chaotic quantum systems is a striking
problem of quantum chaos. In the semiclassical limit quantum chaotic
systems display properties similar to the classical chaos. For
example, decay rate of the quantum  Loschmidt echo is related to the
classical Lyapunov exponent \cite{Jacquod,Iomin,Jal,Cuu}. However
problem of our interest is purely of quantum origin. Since we are
interested in the consequences of quantum chaos in quantum
information processing.

This letter is organized as follows: In Sec.2, we describe the
system and its solution. Sec.3, is devoted to investigate the
dynamics of  entanglement. In Sec.4, The fidelity of the teleported
state is discussed. Finally, we summarize our results in Sec.5.

\section{The Model}

During the last decade many theoretical and experimental efforts have been
done in order to study processes involving atoms inside a cavity, stimulated
by the experimental realization of a multi-photon micromaser \cite{scu97}.
In the rotating wave approximation, the interaction of the cavity mode with
the injected atoms is described by the Hamiltonian \cite{Levan},
\begin{eqnarray}
\mathcal{H}&=&\omega_0(\hat{S_1}^z+\hat{S_2}^z)+\Omega(\hat{S_1}^+{\hat{S_2}%
^-}+ \hat{S_2}^+\hat{S_1}^-)+\omega_f\hat{b}^\dagger \hat{b}  \nonumber \\
&-&g_0\cos(k_f\hat{x})\Bigl((\hat{S_1}^++\hat{S_2}^+)\hat{b}+ (\hat{S_1}^-+%
\hat{S_2}^-)\hat{b}^\dagger\Bigr),
\end{eqnarray}
where, the first term is local spin part, the second represents the
interaction between the two spins, the third is the field cavity
mode Hamiltonian and the last term is the interaction between the
field and the two atoms. The operators $\hat{S}^{\pm}_{i}$ and
$\hat{S}^{z}_{i}, i=1,2$ are the usual raising (lowering) and
inversion operators for the two atoms respectively. Theses operators
satisfy the relations,
\begin{equation}
\Bigl[\hat{S}^{z}_{i},\hat{S}^{\pm}_{i}\Bigr]=\pm 2
\hat{S}^{\pm}_{i}\delta_{ij},\quad
\Bigl[\hat{S}^{+}_{i},\hat{S}^{-}_{i}\Bigr]=\hat{S}^{z}_{i}\delta_{ij},
\end{equation}
where, $\delta_{ij}=1$ if $i=j$ and zero otherwise. The constants
$\omega_0,\omega_f$ are the atoms, the field frequency respectively,
while $\Omega$ represents the coupling constant between the two
spins.

 The term $g_0\cos(k_f \hat{x})$ is coupling between the atoms
and the field, where it depends on the position of the atoms inside
the cavity. This dependence serves as a source of complexity of the
dynamics. As was shown in \cite{Levan}, time dependence $x(t)$
mimics all feature or random Wiener process. The correlation length
of random variable $x(t)$ scales with kinetic energy of the atom on
the input and for very slow initial conditions dynamic is close to
regular. Since the motion of the  adiabatic motion system is
integrable, then the  de-coherence time of the system can be scaled
by proper choice of initial kinetic energy of the atom (see Ref
\cite{Zaslavsky} for example).

In the invariant sub-space of the global system, we can consider a
set of complete basis of the qubit-field system as $\bigl|ee,n-1\bigr\rangle,\bigl|%
eg,n\bigr\rangle,\bigl|ge,n\bigr\rangle$ and $\bigl|gg,n+1\bigr\rangle$.
Assume that the field is initially prepared in a coherent state, $%
\bigl| \psi(0) \bigr\rangle_{f}=\sum_{n=0}^{\infty}{W_n\bigl| n \bigr\rangle}
$. For the atomic system, we consider the initial states of the first and
second atoms are, $\bigl| \psi(0) \bigr\rangle_{1}=a_1\bigl| g \bigr\rangle%
_1+b_1\bigl| e \bigr\rangle_1$, $\bigl| \psi(0) \bigr\rangle_{2}=a_2\bigl| g %
\bigr\rangle_2+b_2\bigl| g \bigr\rangle_2$ respectively , where $1$
stands for the first atom and $2$ for the second atom, with
$|a_i|^2+|b_i|^2=1, i=1, 2$. So, the initial state of the  atomic
system can be written as
\begin{equation}  \label{eeg}
\bigl| \psi_{12}(0) \bigr\rangle=C_{00}\bigl| gg \bigr\rangle+C_{01}\bigl| %
ge \bigr\rangle+C_{10}\bigl| eg \bigr\rangle+C_{11}\bigl| ee
\bigr\rangle),
\nonumber \\
\end{equation}
and consequently the state of the total  system is given by,
\begin{equation}  \label{eegn}
\bigl| \psi_s(0) \bigr\rangle=\sum_{n=0}^{\infty}W_n\Bigr \{\bigl| n %
\bigr\rangle\otimes(C_{00}(0)\bigl| gg \bigr\rangle+C_{01}(0)\bigl| ge %
\bigr\rangle+C_{10}(0)\bigl| eg \bigr\rangle+C_{11}(0) \bigl| ee \bigr\rangle%
)\Bigl\},
\end{equation}
where, $C_{00}(0)=a_1a_2,~ C_{10}(0)=a_1b_2, C_{01}(0)=a_2b_1,
c_{11}(0)=b_1b_2$, $W_{n}=\frac{\bar n ^{n}}{\sqrt{n!}}\exp (-\frac{1}{2}%
|\bar n|^{2})$ and $\bar n$ is the mean photon number inside the
cavity. At any time $t>0$, the state of the atomic system and the
field  is given by
\begin{equation}
\bigl| \psi_s(t) \bigr\rangle=\sum_{n=0}^{\infty}\Bigl\{ A_n(t)\bigl| gg,n+1 %
\bigr\rangle+B_n(t)\bigl| ge,n \bigr\rangle+C_n(t)\bigl| eg,n \bigr\rangle%
+D_n(t) \bigl| ee,n-1 \bigr\rangle\bigr\},
\end{equation}
where,
\begin{eqnarray}
A_{n}(t) &=&W_{n}\Bigl\{\frac{\sqrt{n+1}}{2\sqrt{2}{(2n+1)}}e^{-i\Omega
t}\left( C_{00}(0)Q[\omega (t)]-C_{01}(0)Q^{-1}[\omega (t)]\right)
\nonumber \\
&&+\frac{e^{-i\Omega t}}{2\sqrt{n+1}}\Bigl(C_{11}(0)-\frac{1}{2\sqrt{2}(2n+1)%
}(C_{00}(0)-C_{01}(0))\Bigr)\Bigr\},  \nonumber \\
B_{n}(t) &=&W_{n}\Bigl\{\frac{e^{-i\Omega t}}{4\sqrt{2n+1}}%
(C_{00}(0)Q[\omega (t)]-C_{01}(0)Q^{-1}[\omega (t)])-\frac{C_{10}(0)}{2}%
e^{i\Omega t}\Bigr\},  \nonumber \\
C_{n}(t) &=&W_{n}\Bigr\{\frac{e^{-i\Omega t}}{4\sqrt{2n+1}}%
(C_{00}(0)Q[\omega (t)]-C_{01}(0)Q^{-1}[\omega (t)])+\frac{C_{10}(0)}{2}%
e^{i\Omega t}\Bigr\},  \nonumber \\
D_{n}(t) &=&W_{n}\Bigl\{\frac{\sqrt{n}}{2\sqrt{2}{(2n+1)}}e^{-i\Omega
t}\left( C_{00}(0)Q[\omega (t)]-C_{01}(0)Q^{-1}[\omega (t)]\right)
\nonumber \\
&&-\frac{e^{-i\Omega t}}{2\sqrt{n+1}}\Bigl(C_{11}(0)-\frac{1}{2\sqrt{2}(2n+1)%
}(C_{01}(0)-C_{00}(0))\Bigr)\Bigr\},
\end{eqnarray}%
and
\begin{equation}
Q[\omega (t)]=e^{i\int_{0}^{1}{\omega {\grave{(t})}d\grave{t}}}~\mbox{with}%
~\omega (t)=\sqrt{2(2n+1)}g_{0}\cos (k_{f}x(t)),~Q^{-1}[\omega (t)]=Q^{\ast
}[\omega (t)].
\end{equation}%
Due to the chaotic motion of atoms inside of cavity, we consider the
random parameter, $x(t)$ as a classical chaotic function. Therefore
 $Q[\omega (t)]$ should be averaged over all
possible realizations of the  random parameter $\omega (t)$. In
this treatment,  we consider  the case of weak chaos,
corresponding to the large values of correlation time $\tau _{c}$
of random variable $x(t)$ \cite{Levan}.

 Since we are interested in
discussing some properties of the atomic system, we calculate the
density matrix of the two atoms by tracing out the field
i.e $\varrho_{12}=tr_{f}\{\varrho_{Af}\}$, where $\varrho_{Af}=\bigl| %
\psi_s(t) \bigr\rangle\bigl\langle \psi_s(t) \bigr|$ and $\varrho_{12}$ is
the density operator of the atomic system. In an explicit form,
\begin{eqnarray}\label{atomic}
\varrho_{12}(t)&=&\bigl| gg \bigr\rangle\sum_{n=0}^{\infty}\Bigl(| A_n|^2%
\bigl\langle gg \bigr|+ A_n B^*_{n+1}\bigl\langle ge \bigr|+ A_n C^*_{n+1}%
\bigl\langle eg \bigr|+ A_n D^*_{n+2}\bigl\langle ee \bigr|\Bigr)  \nonumber
\\
&&+\bigl| ge \bigr\rangle\sum_{n=0}^{\infty}\Bigl( B_n A^*_{n-1}\bigl\langle %
gg \bigr|+|B_n|^2\bigl\langle ge \bigr|+ B_n C^*_n\bigl\langle eg \bigr|+
B_n D^*_{n+1}\bigl\langle ee \bigr|\Bigr)  \nonumber \\
&& +\bigl| eg \bigr\rangle\sum_{n=0}^{\infty}\Bigl( C_n A^*_{n-1}%
\bigl\langle gg \bigr|+ C_n B^*_n\bigl\langle ge \bigr| +| C_n|^2%
\bigl\langle eg \bigr|+ C_n D^*_{n+1}\bigl\langle ee \bigr|\Bigr)  \nonumber
\\
&&+\bigl| ee \bigr\rangle\sum_{n=0}^{\infty}\Bigl( D_n A^*_{n-2}\bigl\langle %
gg \bigr|+ D_n B^*_{n-1}\bigl\langle ge \bigr|+ D_n C^*_n\bigl\langle eg %
\bigr|+| D_n|^2\bigl\langle ee \bigr|\Bigr).
\end{eqnarray}
Having the density operator, we can investigate all the physical
properties of the  atomic system.
\begin{figure}[tbp]
\begin{center}
\includegraphics[width=19pc,height=12pc]{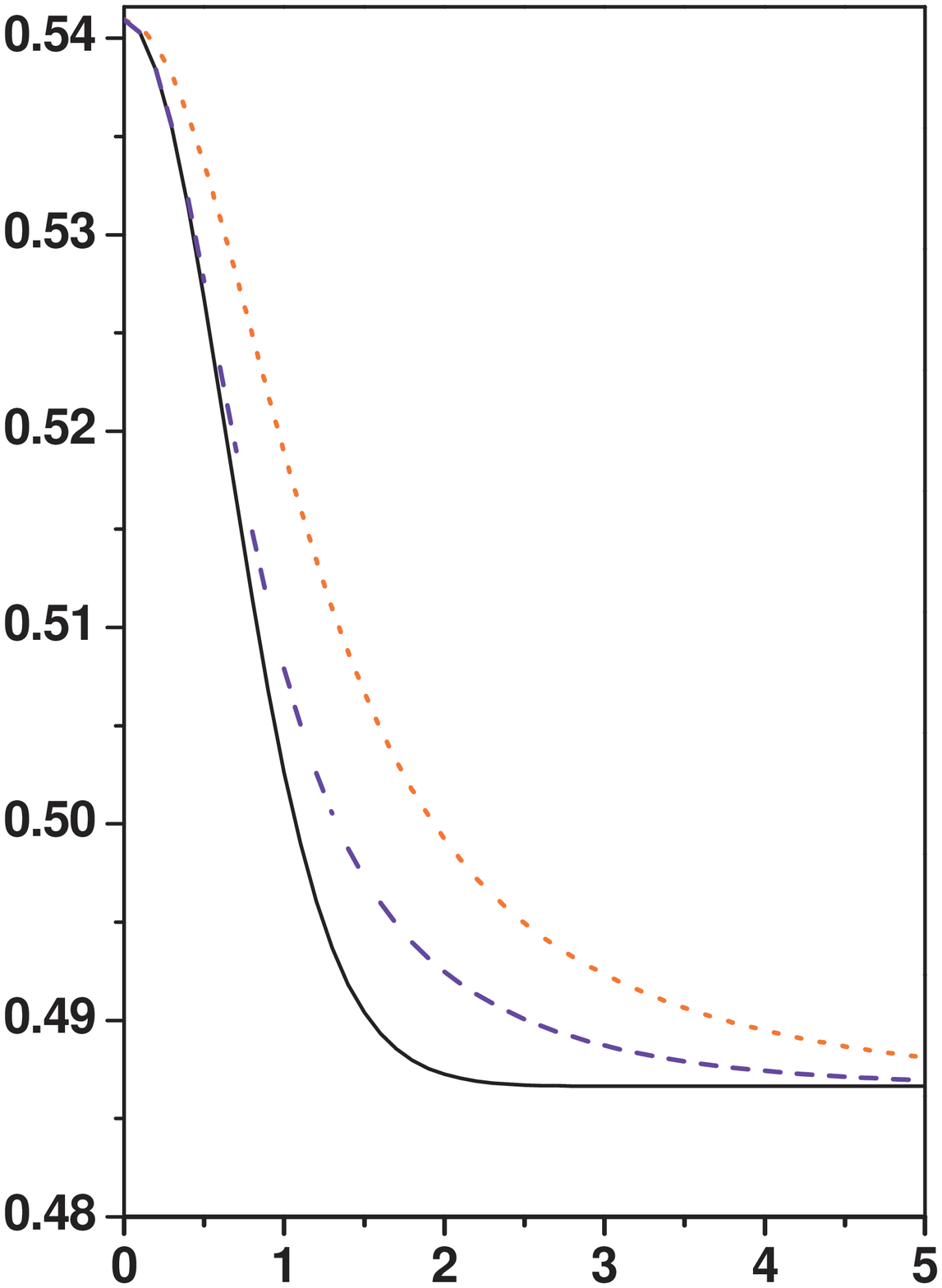}
  \put(-110,10){t}
  \put(-50,115){$(a)$}
   \includegraphics[width=18pc,height=12pc]{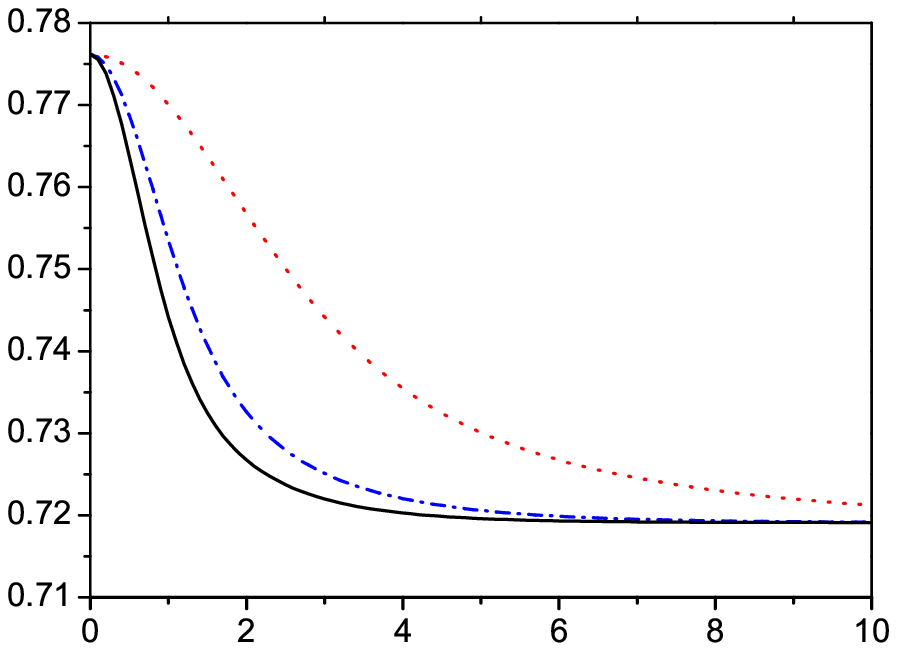}  \put(-100,10){t}
\put(-450,70){DoE} \put(-230,70){DoE}
  \put(-100,10){t}
   \put(-35,115){$(b)$}
\end{center}
\caption{The degree of entanglement between two atoms initially
prepared in the state (\ref{eeg}) with $C_{00}=0.2, C_{01}=C_{10}=0$
and $C_{11}=\sqrt{1-C_{00}^2}$. The chaotic parameter,
$\protect\gamma=0.1,0.5$ and $0.9$, for the solid dash and dot
curves respectively (a)$\bar n=5$ (b)$\bar n=6$.}
\end{figure}

\section{Entanglement}
To quantify the amount of entanglement contained in the entangled
states, we shall use a measurement called negativity  introduced
first by K. Zyczkowski \cite{Zyc}. This measure states that if the
eigenvalues of the partial transpose are given by $\mu_i; i = 1, 2,
3, 4,$ then the degree of entanglement, DOE is given by,
\begin{equation}  \label{DoE}
DoE=\sum_{i=1}^{4}{\mu_i}-1,
\end{equation}
where, $\mu_i$ are the eigenvalues of the partial transpose of the density
operator $\rho_{12}^{^{\mathsf{T}_{\!2}}}(t) $.

To see the effect of the random parameter on the dynamics of
entanglement, we should average the random funstion
 $Q[\omega(t)]$ over all possible realizations of $x(t)$  as
 \cite{Levan},
\begin{equation}
\expect{Q[\omega(t)]}= exp[-\frac{t}{2}\sqrt{\pi\gamma}
Erf(t\sqrt{\gamma})].
\end{equation}
Now, investigating the effect of the function $x(t)$ on the dynamics of
entanglement is equivalent to investigate the effect of the parameter $%
\gamma$.

Fig.(1), shows the dynamics of entanglement for different values of
the chaotic  parameter $\gamma$, where we assume that the atomic
system is initially prepared in a partial entangled state
$\varrho_{12}=0.2\ket{gg}\bra{gg}+\sqrt{0.6}\ket{ee}\bra{ee}$. In
Fig.(1a), we consider the mean photon numbers inside the cavity
$\bar n=5$.  it is clear that, for small values of the parameter
$\gamma$, the entanglement decays smoothly. However, for large
values of $\gamma$, the DOE, decays faster and then become stable.
This means that even if  the system suffers from the chaotic
environment the entangle stated is robust.  In Fig.(1b), describes
the dynamics of entanglement for $\bar n=6$.  One can notice that
the behavior of the entanglement is the same as that depicted in
Fig.(1a), namely,  the entanglement decays  fast as one increases
the chaotic parameter $\gamma$. From Fig.($1a\&1b$), it is clear
that, as one increases the $\bar n$ inside the cavity, the
entanglement improved. This behavior of entanglement is  similar to
that depicted in \cite{Fur}, where the entropy increases as the time
increases. Also,  due to the damping effect  the oscillations are
very small and appears only for a long range of time. This explain
the disappearance of oscillation behavior of the entanglement.

So, one concludes then, the entanglement decays faster as one
increases the chaotic parameter. Also, this effect does not cause
a death of entanglement. This means  that,  the traveling partial
entangled state in the chaotic cavity is robust. However, by
increasing the mean photon numbers inside the cavity, one can
delay the decay and enhance the degree of entanglement.

\section{Teleportation}

Teleportation is one of the most promising candidate in quantum
communation theory \cite{Niel}. Since, it has been discovered by
Bennett et. al \cite{ben}, there are different protocols have been
developed in different directions. Performing quantum
teleportation by using entangled atoms has been investigated by
many authors\cite{Metwally,Rie,Mic}. One of the most practical
application is quantum teleportation through noise channels( see
for example\cite{HWa,Eylee,Sah, Nasser}).

In this study, we investigate the effect of the chaotic parameter
$\gamma$, on the fidelity of the teleported state.  Let us assume
that we have two users,  Alice and Bob share an entangled state
given by (\ref{atomic}). Alice is  given unknown state
$\rho_u=\bigl| \psi_u \bigr\rangle\bigl\langle \psi_u \bigr|$
where,
\begin{equation}\label{ru}
\bigl| \psi_u \bigr\rangle=\alpha\bigl| g \bigr\rangle+\beta\bigl| e %
\bigr\rangle \mbox{ with}\quad |\alpha|^2+|\beta|^2=1.
\end{equation}
The aim of Alice is sending this unknown state to Bob, where they
use the original teleportation protocol \cite{ben}.

To achieve this task, the partners follow the following steps:
\begin{enumerate}
\item Alice performs the  CNOT gate on her qubit and the given
unknown qubit  followed by  Hadamard gate.

\item Alice measures her qubit and the unknown qubit randomly in
one of the basis
$\ket{\phi^\pm}=\frac{1}{\sqrt{2}}(\ket{ee}\pm{\ket{gg}}),
\ket{\psi^\pm}=\frac{1}{\sqrt{2}}(\ket{eg}\pm{\ket{ge}})$,
 and sends her results to Bob by using classical
channel.

\item As soon as Bob receives the classical data from Alice, he
applies a single qubit operation on his qubit depending on Alice's
results.
\end{enumerate}
If, Alice measures  the Bell state, $\bigl|\phi ^{+}\bigr\rangle=\frac{1}{%
\sqrt{2}}(\bigl| ee \bigr\rangle+\bigl| gg \bigr\rangle)$, then the density
operator on Bob's hand, $\rho_b$, is given by,
\begin{equation}\label{r-Bob}
\rho_{b}=\kappa_1\bigl| g \bigr\rangle\bigl\langle g \bigr|+\kappa_2\bigl| g %
\bigr\rangle\bigl\langle e \bigr|+ \kappa_3\bigl| e \bigr\rangle\bigl\langle %
g \bigr|+\kappa_4\bigl| e \bigr\rangle\bigl\langle e \bigr|,
\end{equation}
where,
\begin{eqnarray}
\kappa_1&=&\frac{1}{2}\sum_{n=0}^{\infty}\Bigl\{|\alpha|^2|A_n|^2+
\alpha\beta^*A_nC^*_{n+1}+ \beta\alpha^*C_nA_{n-1}^*+|\beta|^2|C_n|^2 \Bigr\}%
,  \nonumber \\
\kappa_2&=&\frac{1}{2}\sum_{n=0}^{\infty}\Bigl\{|\alpha|^2A_nB^*_{n+1}+
\alpha\beta^*A_nD^*_{n+2}+\beta\alpha^* c_n B^*_n+|\beta|^2C_nD^*_{n+2}%
\Bigr\},  \nonumber \\
\kappa_3&=&\frac{1}{2}\sum_{n=0}^{\infty}\Bigl\{(|\alpha|^2B_nA^*_{n-1}+
\alpha\beta^*B_nC^*_{n}+\beta\alpha^*D_n A^*_{n-2}+ |\beta|^2D_nC^*_{n}%
\Bigr\},  \nonumber \\
\kappa_4&=&\frac{1}{2}\sum_{n=0}^{\infty}\Bigl\{(|\alpha|^2|B_n|^2+
\alpha\beta^*B_nD^*_{n+1}+\beta\alpha^*D_n B^*_{n-1}+|\beta|^2|D_n|^2\Bigr\}.
\end{eqnarray}
To quantify the closeness of the initial state (\ref{ru}) with the
final state (\ref{r-Bob}), we evaluate the fidelity $\mathcal{F}$
which is given by,
\begin{equation}\label{fid}
\mathcal{F}=tr\Bigl\{\rho_u\rho_b\Bigr\}.
\end{equation}

The dynamics of the fidelity  $\mathcal{F}$, of the teleported
state (\ref{ru}) is presented in Fig.(2). This figure displays the
possibility of using the state $\varrho_{12}$ as a quantum channel
for a short time to implement the quantum teleportation protocol.
 It is clear that, in the case of ideal environment namely
 $\gamma=0$,  Alice can send the unknown state (\ref{ru}) with high
 fidelity. However for small value of the chaotic parameter
 $\gamma$, the fidelity $\mathcal{F}$ decreases. As one increases
 the chaotic parameter more, one can see that the fidelity decreases
 fast and at $\gamma=1$, the fidelity is zero. This behavior due to
 the lose of entanglement as one increases the chaotic parameter

A visible behavior of the fidelity is given in Fig.(3), where it
describes $\mathcal{F}$ as a contour lines. From this figure, we
can determine the interval of  the critical time $\tau_c$, after
which the effect of the chaotic parameter appears. It is easily
seen that, the teleported state   can be transmitted with high
fidelity from one location to another for any value of the chaotic
parameter $\gamma$ in the interval $t \in[0,0.25]$. However, for
small values of $\gamma \in[0,0.14]$, one can teleporte the
information safely with $\mathcal{F}=1$.

For $t>0.25$, the fidelity decays smoothly and sharply for larger
time. As one increases $\gamma$ more, the fidelity decreases
smoothly and then becomes constant. This behavior is seen for any
value of the chaotic parameter $\gamma$. Also, one can see that as
one increases the chaotic parameter, the dark sectors increases.
This means that the fidelity decreases. However the bright regions
increases as one decreases the value of the chaotic parameter.

\begin{figure}[tbp]
\begin{center}
\includegraphics[width=20pc,height=15pc]{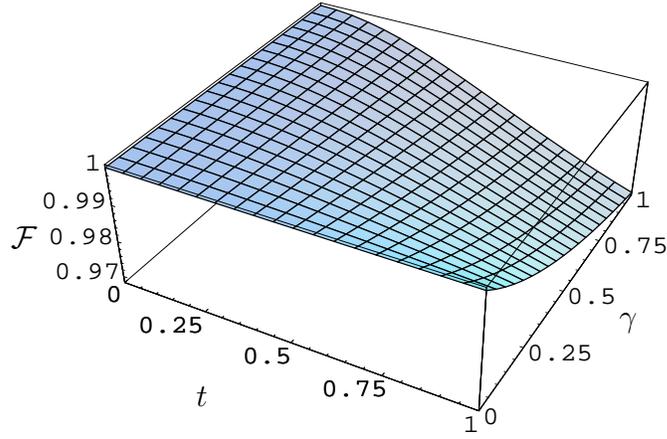} \put(-180,20){$t$}
\put(-250,80){$\mathcal{F}$} \put(-20,50){$\gamma$}
\end{center}
\caption{ The Fidelity of the tleported sate with $\protect\alpha=0.95$ and $%
\protect\beta=\protect\sqrt{1-\protect\alpha^2}$ The initial state is
prepared such that $C_{00}=0.2, C_{01}=C_{10}=0, C_{11}=\protect\sqrt{%
1-C_{00}^2}$ and $\bar n=5, \Omega=1$}
\end{figure}
\begin{figure}[tbp]
\begin{center}
\includegraphics[width=20pc,height=15pc]{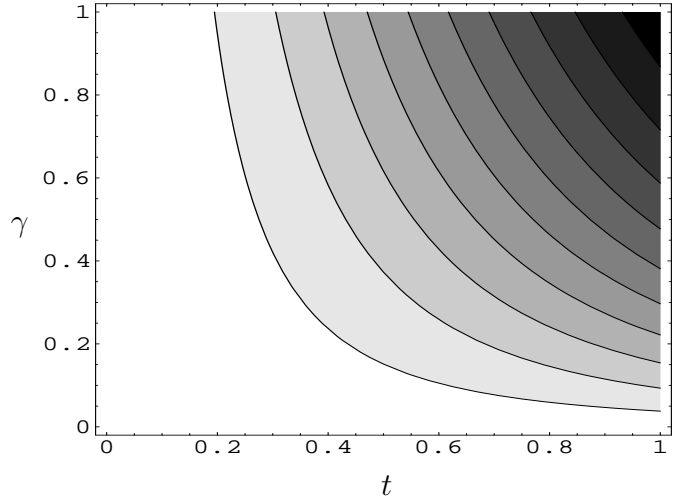}
\put(-110,-10){$t$}  \put(-250,90){$\gamma$} %
\end{center}
\caption{ The fidelity, $\mathcal{F}$ of the teleported as contour
line against the time $ t$  and the chaotic parameter
$\protect\gamma$. }
\end{figure}

From theses results, one can say that it is possible to  achieve
quantum teleportation for any value of the chaotic  parameter with
reasonable fidelity. Also, for small range of time  one can
teleported the information with unit fidelity for any  value of
the chaotic parameter.

\section{Conclusion}

Usually sudden death of Entanglement is associated  with the
environmental effects. It could be radiation field or thermal
fluctuations as well. Question what kind consequences nonuniform
cavity may have on the quantum information processing was missed
yet. Idea of nonuniform cavity (which is of course more realistic
from experimental point of view than uniform cavity) implies that
constant of coupling between radiation field and optical atom
depends on atoms position inside of cavity. This dependence is as a
rule ignored for standard Jaynes-Cummings model.

However the nonuniform cavity may lead to the complex dynamic of the
system.  The aim of the present paper is investigating  the
consequences of the nonuniform cavity on the quantum information
processing. With this in the aim, we studied particular model system
to identify new mechanism of decoherence. Key issue is that in our
case, in contrast to the common scenario (thermal and environmental
effects), sours of decoherence  is the initial random parameter over
which averaging procedure is done.  This parameter itself is a
consequence of the random motion of atoms due to nonuniform quantum
cavity. Interest to this problem is motivated by the fact that in
case of low temperatures, and in the cavities with small mean photon
numbers, nonuniform effects on the quantum information processing
becomes important, since it recovers thermal and environmental
effects.

Basically effect of the random parameter on the degree of
entanglement for a two-qubit system and on the fidelity of the
teleported state is investigated. The time behavior for both
phenomenon, is depicted for the time interval comparable with the
correlation length of random parameter. We observe decay of degree
of entanglement  with time if underling dynamic is chaotic. Decay
rate is defined by  features of atomic motion inside of cavity and
may be scaled along with the correlation length of random variable.
Advantage of the offered model is that, decay rate is quite low if
initial kinetic energy of the system is small.
 Also, an important result is that, in spite of de-coherence,
 minimum degree of entanglement does not reach to zero.
This indicates on the possibility, of using given model for
performing quantum teleportation with reasonable fidelity, during the large lapse of time.

\textbf{Acknowledgment:} We are grateful to the referee for their
constructive comments and remarks have improve our results. The
designated project has been fulfilled by financial support from the
Georgian National Foundation (grants: GNSF/STO 7/4-197, GNSF/STO
7/4-179). The financial support of Deutsche Forschungsgemeinschaft
through SPP 1285 (contract number EC94/5-1) is gratefully
acknowledged by L. Chotorlishvili.

\[
\]

\end{document}